\documentstyle[vsolj01,graphicx,natbib]{article}

\RequirePackage[T1]{fontenc}

\def\cite{\citealt}

\begin{document}

\title{ASASSN-V J205543.90$+$240033.5: another white dwarf pulsar?}

\author{Taichi Kato$^1$}
\author{$^1$ Department of Astronomy, Kyoto University,
       Sakyo-ku, Kyoto 606-8502, Japan}
\email{tkato@kusastro.kyoto-u.ac.jp}

\begin{abstract}
I found that ASASSN-V J205543.90$+$240033.5 shows
large-amplitude (1.2--1.4~mag) nearly sinusoidal variations
with a period of 10.803(2)~d and very short period variations
with a period of 0.0068~d using
Public Data Release of Zwicky Transient Facility
observations.  The only known object that shows a similar
combination of nearly sinusoidal reflection variations and
very short period, large-amplitude variations
is the unique white dwarf pulsar AR Sco.
ASASSN-V J205543.90$+$240033.5 appears to be very similar
to AR Sco and observations at various wavelengths
are desired.
\end{abstract}

   ASASSN-V J205543.90$+$240033.5 is variable object detected
by the All-Sky Automated Survey for Supernovae (ASAS-SN,
\cite{ASASSN}, \cite{koc17ASASSNLC}) and was cataloged as
and eclipsing binary (type EB) with a period of 1.046984~d.\footnote{
Variable Stars Database (AVSD)
$<$http://asas-sn.osu.edu/variables$>$ on 23-Oct-2020)
(22-Oct-2019 version).
}
This object was also independently listed as a candidate
RR Lyr star by \citet{ses17PS1RR} using the Panoramic Survey
Telescope and Rapid Response System (Pan-STARRS1, \cite{PS1}).
It was also independently listed as a variable star
ATO J313.9329$+$24.0092 by \citet{hei18ATLASvar}
using the Asteroid Terrestrial-impact Last Alert System (ATLAS,
\cite{ATLAS}) data.

\begin{figure*}
  \begin{center}
    \includegraphics[width=16cm]{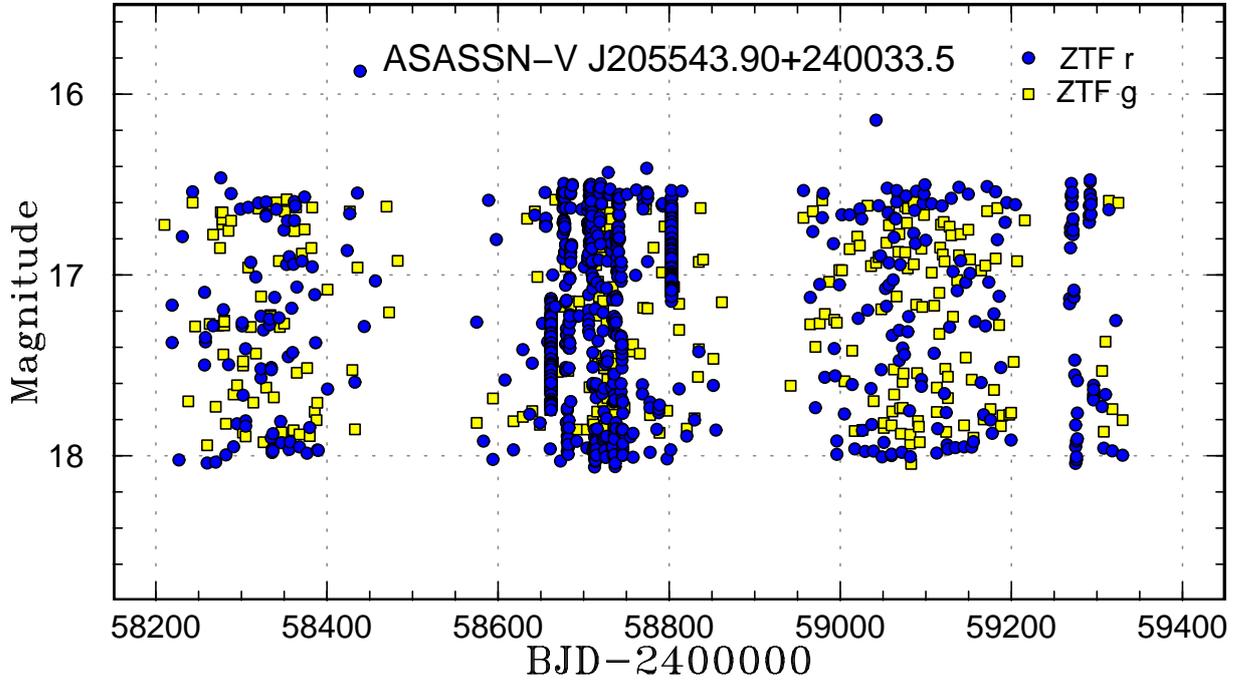}
  \end{center}
  \caption{Long-term variations of ASASSN-V J205543.90$+$240033.5
  in the ZTF data.}
  \label{fig:lcall}
\end{figure*}

\begin{figure*}
  \begin{center}
    \includegraphics[width=16cm]{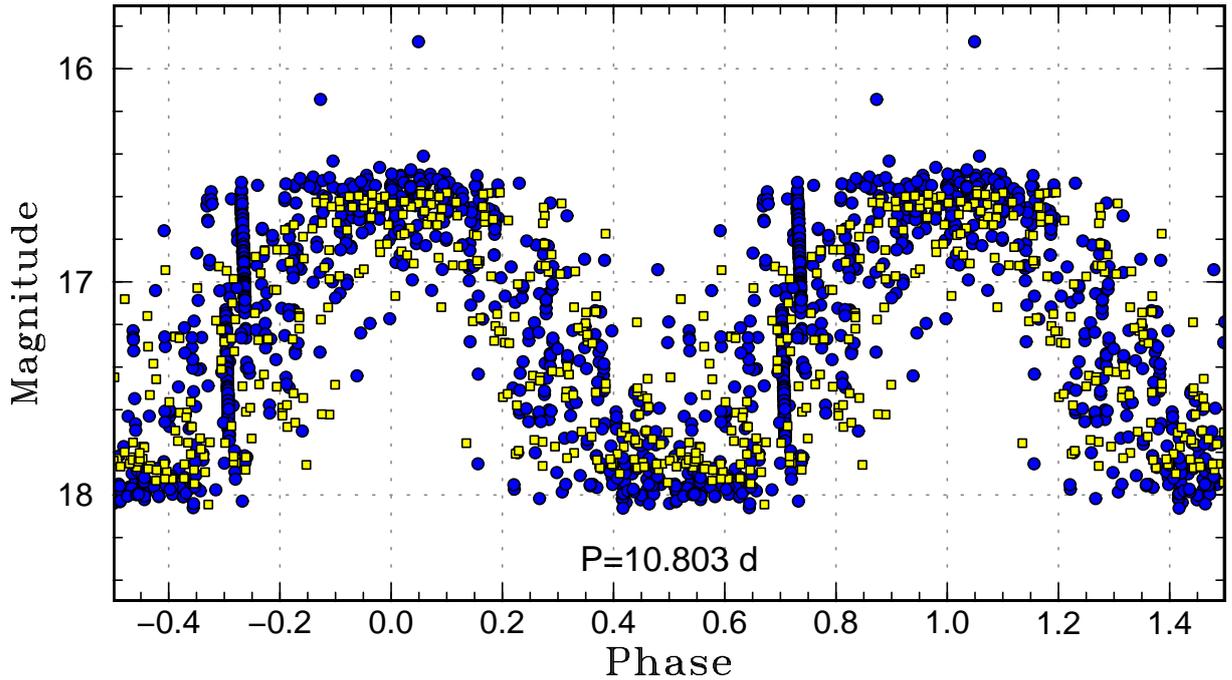}
  \end{center}
  \caption{Phase-folded variations of ASASSN-V J205543.90$+$240033.5
  in the ZTF data.  The symbols are the same as in
  figure \ref{fig:lcall}.  The epoch was chosen as BJD 2458773.10.}
  \label{fig:phorb}
\end{figure*}

   I used Public Data Release 6 of
the Zwicky Transient Facility \citep{ZTF}
observations\footnote{
   The ZTF data can be obtained from IRSA
$<$https://irsa.ipac.caltech.edu/Missions/ztf.html$>$
using the interface
$<$https://irsa.ipac.caltech.edu/docs/program\_interface/ztf\_api.html$>$
or using a wrapper of the above IRSA API
$<$https://github.com/MickaelRigault/ztfquery$>$.
} and found that this object showed large-amplitude
(1.2--1.4~mag) nearly sinusoidal variations with a period
of 10.803(2)~d and very short period variations
during time-resolved runs on two nights in the ZTF data
(figures \ref{fig:lcall}, \ref{fig:phorb}).

   The object is very blue as can be seen as the almost
zero color index in figures \ref{fig:lcall}, \ref{fig:phorb}.
The large distance modulus (11.1~mag) based on 
the Gaia parallax \citep{GaiaEDR3} indicates that
the object is intrinsically faint.  These features suggest
that this 10.803-d day variation is caused by reflection by 
the secondary star in a binary containing a hot component.

\begin{figure*}
  \begin{center}
    \includegraphics[width=16cm]{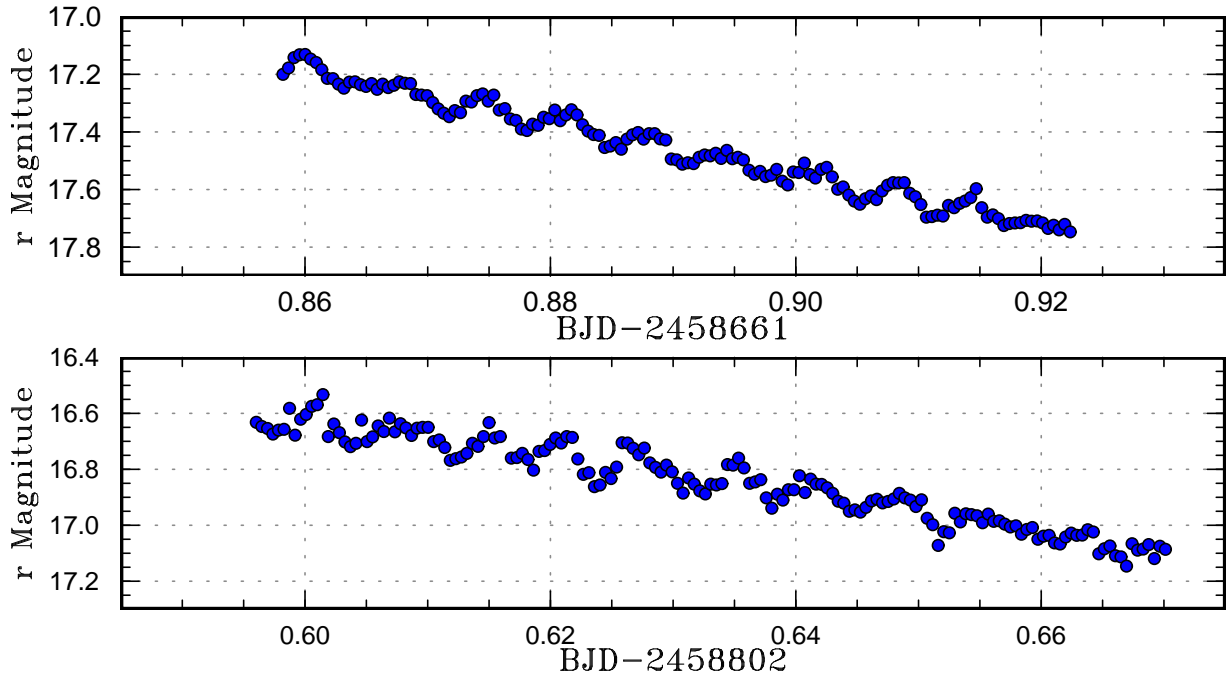}
  \end{center}
  \caption{Short-term variations of ASASSN-V J205543.90$+$240033.5
  in the ZTF data.}
  \label{fig:short}
\end{figure*}

\begin{figure*}
  \begin{center}
    \includegraphics[width=16cm]{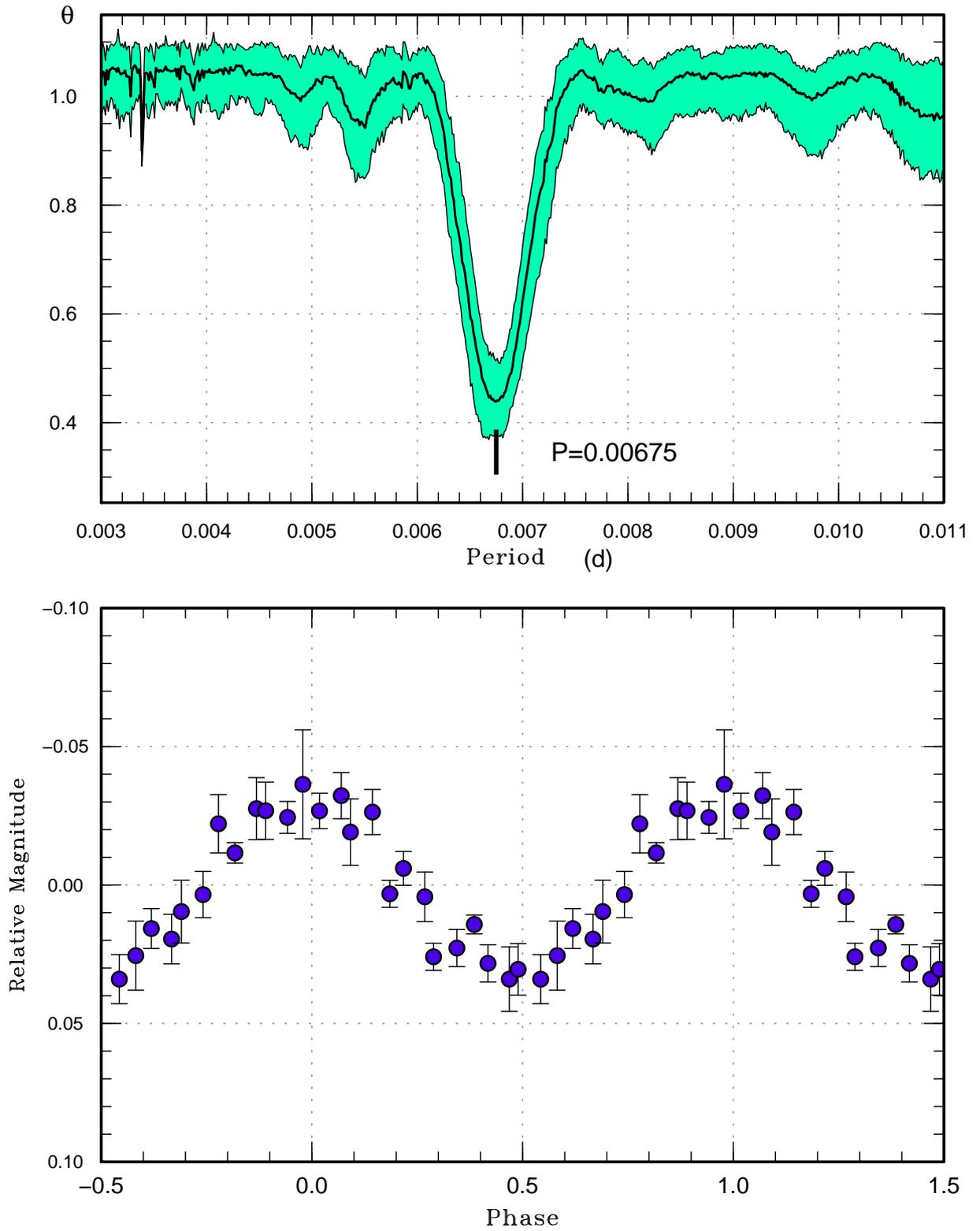}
  \end{center}
  \caption{Pulse profile of ASASSN-V J205543.90$+$240033.5
    on BJD 2458661.
    (Upper): PDM analysis.
    We analyzed 100 samples which randomly contain 50\% of
    observations, and performed the PDM analysis for these samples.
    The bootstrap result is shown as a form of 90\% confidence intervals
    in the resultant PDM $\theta$ statistics.
    (Lower): Phase-averaged profile.
    }
  \label{fig:pulse1}
\end{figure*}

\begin{figure*}
  \begin{center}
    \includegraphics[width=16cm]{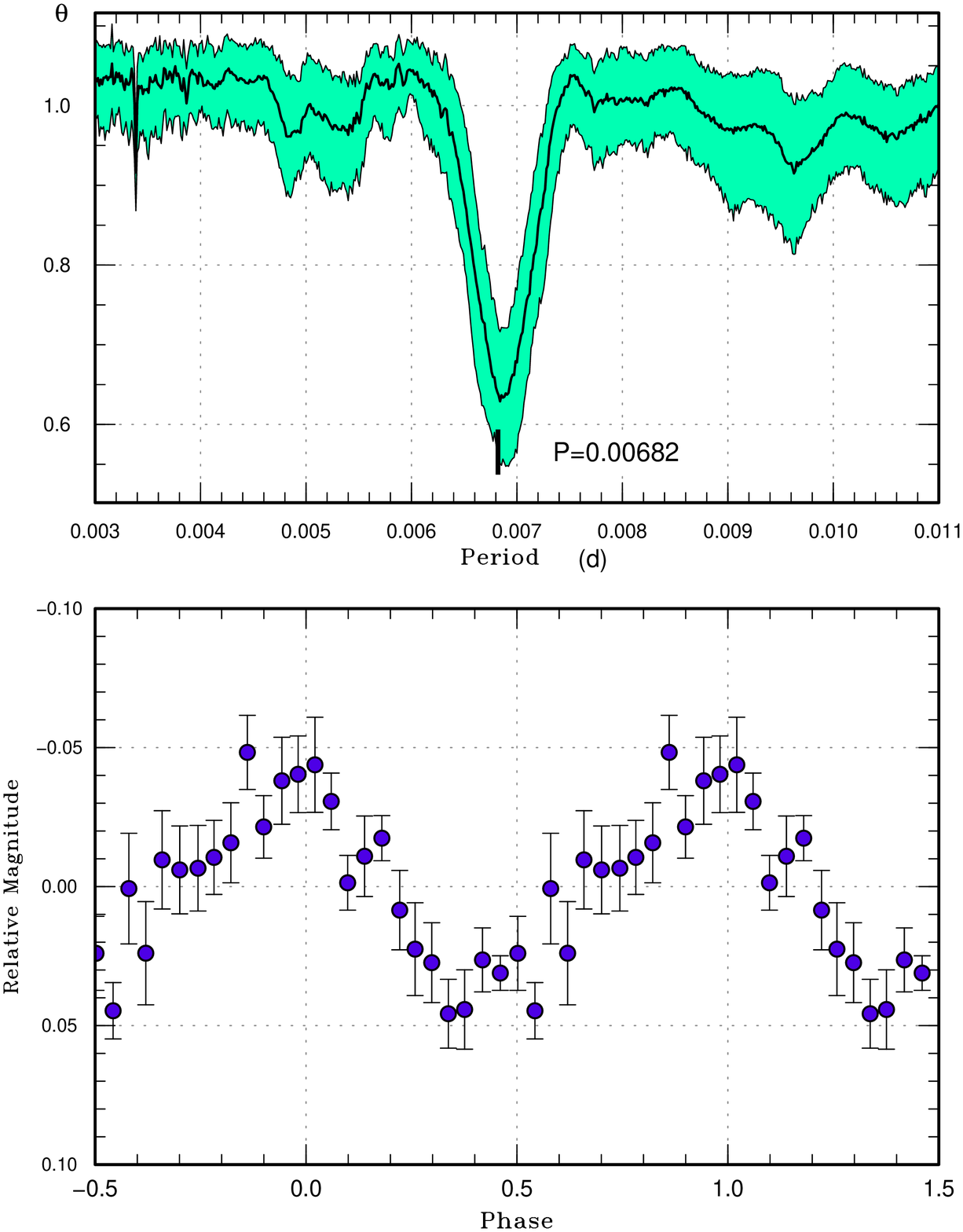}
  \end{center}
  \caption{Pulse profile of ASASSN-V J205543.90$+$240033.5
    on BJD 2458802.}
  \label{fig:pulse2}
\end{figure*}

   The short-term variations are directly visible to
the eye in the ZTF time-resolved photometry
(figure \ref{fig:short}).  Phase Dispersion Minimization
(PDM; \cite{PDM}) analysis after de-trending the data
yielded periods of 0.00675(3)~d for the BJD 2458661 run
(figure \ref{fig:pulse1}) and 0.00682(2)~d for
the BJD 2458802 run (figure \ref{fig:pulse2}).
The errors of the periods were determined by
the methods of \citet{fer89error} and \citet{Pdot2}.
These periods are fairly in agreement and I consider
that they represent the spin period of the white dwarf.

   This combination of the large-amplitude reflection
variation and the high-amplitude spin pulse is seen
in the white dwarf pulsar AR Sco
(\cite{mar16arsco}; \cite{sti18arsco}).
We recently reported that ZTF J185139.81$+$171430.3 =
ZTF18abnbzvx shows high-amplitude spin variations
\citep{kat21j185139} and suggested it to be
a possible white dwarf pulsar, but the large-amplitude
reflection variation is missing in ZTF J185139.81$+$171430.3.
The present observations indicate ASASSN-V J205543.90$+$240033.5
looks more similar to AR Sco than ZTF J185139.81$+$171430.3 and
observations in various wavelengths to search for the evidence
for non-thermal emissions as in AR Sco.

\section*{Acknowledgments}

The author is grateful to Naoto Kojiguchi for supplying
a wrapper code for obtaining the ZTF data.

This work was supported by JSPS KAKENHI Grant Number 21K03616.

Based on observations obtained with the Samuel Oschin 48-inch
Telescope at the Palomar Observatory as part of
the Zwicky Transient Facility project. ZTF is supported by
the National Science Foundation under Grant No. AST-1440341
and a collaboration including Caltech, IPAC, 
the Weizmann Institute for Science, the Oskar Klein Center
at Stockholm University, the University of Maryland,
the University of Washington, Deutsches Elektronen-Synchrotron
and Humboldt University, Los Alamos National Laboratories, 
the TANGO Consortium of Taiwan, the University of 
Wisconsin at Milwaukee, and Lawrence Berkeley National Laboratories.
Operations are conducted by COO, IPAC, and UW.

The ztfquery code was funded by the European Research Council
(ERC) under the European Union's Horizon 2020 research and 
innovation programme (grant agreement n$^{\circ}$759194
-- USNAC, PI: Rigault).

\end{document}